\documentclass[aps,prx,twocolumn,superscriptaddress,floatfix,longbibliography,graphicx]{revtex4-2}
\usepackage{graphicx}
\usepackage{dcolumn}
\usepackage{bm}
\usepackage{gensymb}
\usepackage{amsmath} 
\usepackage{amssymb}
\usepackage{multirow}
\usepackage{xr}
\usepackage{xcolor}
\usepackage{float}
\usepackage{soul}

\begin{document}
\title{Multiple lattice instabilities and complex ground state in Cs$_2$AgBiBr$_6$}

\author{Xing He}
\affiliation{Mechanical Engineering and Materials Science, Duke University, Durham, NC, USA \\}

\author{Matthew Krogstad}
\affiliation{Materials Science Division, Argonne National Laboratory, Lemont, IL, USA \\}
\affiliation{X-ray Sciences Division, Argonne National Laboratory, Lemont, IL, USA \\}

\author{Mayanak K Gupta} 
\affiliation{Mechanical Engineering and Materials Science, Duke University, Durham, NC, USA \\}
\affiliation{Solid State Physics Division, Bhabha Atomic Research Centre, Mumbai 400085, India\\}

\author{Tyson Lanigan-Atkins}
\affiliation{Mechanical Engineering and Materials Science, Duke University, Durham, NC, USA \\}

\author{Chengjie Mao}
\affiliation{Mechanical Engineering and Materials Science, Duke University, Durham, NC, USA \\}

\author{Feng Ye}
\affiliation{Neutron Scattering Division, Oak Ridge National Laboratory, Oak Ridge, TN, USA \\}

\author{Yaohua Liu}
\affiliation{Neutron Scattering Division, Oak Ridge National Laboratory, Oak Ridge, TN, USA \\}

\author{Tao Hong}
\affiliation{Neutron Scattering Division, Oak Ridge National Laboratory, Oak Ridge, TN, USA \\}

\author{Songxue Chi}
\affiliation{Neutron Scattering Division, Oak Ridge National Laboratory, Oak Ridge, TN, USA \\}

\author{Haotong Wei}
\affiliation{Applied Physical Sciences, University of North Carolina at Chapel Hill, Chapel Hill, NC, USA\\}

\author{Jinsong Huang}
\affiliation{Applied Physical Sciences, University of North Carolina at Chapel Hill, Chapel Hill, NC, USA\\}

\author{Stephan Rosenkranz}
\affiliation{Materials Science Division, Argonne National Laboratory, Lemont, IL, USA \\}

\author{Raymond Osborn} 
\affiliation{Materials Science Division, Argonne National Laboratory, Lemont, IL, USA \\}

\author{Olivier Delaire}
\email{olivier.delaire@duke.edu}
\affiliation{Mechanical Engineering and Materials Science, Duke University, Durham, NC, USA \\}
\affiliation{Department of Physics, Duke University, Durham, NC, USA \\}
\affiliation{Department of Chemistry, Duke University, Durham, NC, USA \\}

\date{\today}

\begin{abstract}
Metal halides perovskites (MHPs) are attracting considerable interest for optoelectronic applications, with Cs$_2$AgBiBr$_6$ one of the main contenders among lead-free systems. 
Cs$_2$AgBiBr$_6$ crystallizes in a nominally double-perovskite structure, but exhibits a soft lattice with large atomic fluctuations characteristic of MHPs. While crucial to understand electron-phonon and phonon-phonon couplings, the spatio-temporal correlations of these fluctuations remain largely unknown.
Here, we reveal these correlations using comprehensive neutron and x-ray scattering measurements on Cs$_2$AgBiBr$_6$ single-crystals, complemented with first-principles simulations augmented with machine-learned neural-network potentials. 
We report the discovery of an unexpected complex modulated ground state structure containing several hundred atoms, arising from a soft-phonon instability of the low-temperature tetragonal phase. 
Further, our experiments and simulations both reveal extensive correlated 2D fluctuations of Br octahedra at finite temperature, arising from soft anharmonic optic phonons, reflecting very shallow potential wells.
These results provide new insights into the atomic structure and fluctuations in MHPs, critical to understand and control their thermal and optoelectronic properties.
\end{abstract}
\maketitle

\section{Introduction}

\begin{figure*}
	\centering
	\includegraphics[width=\textwidth]{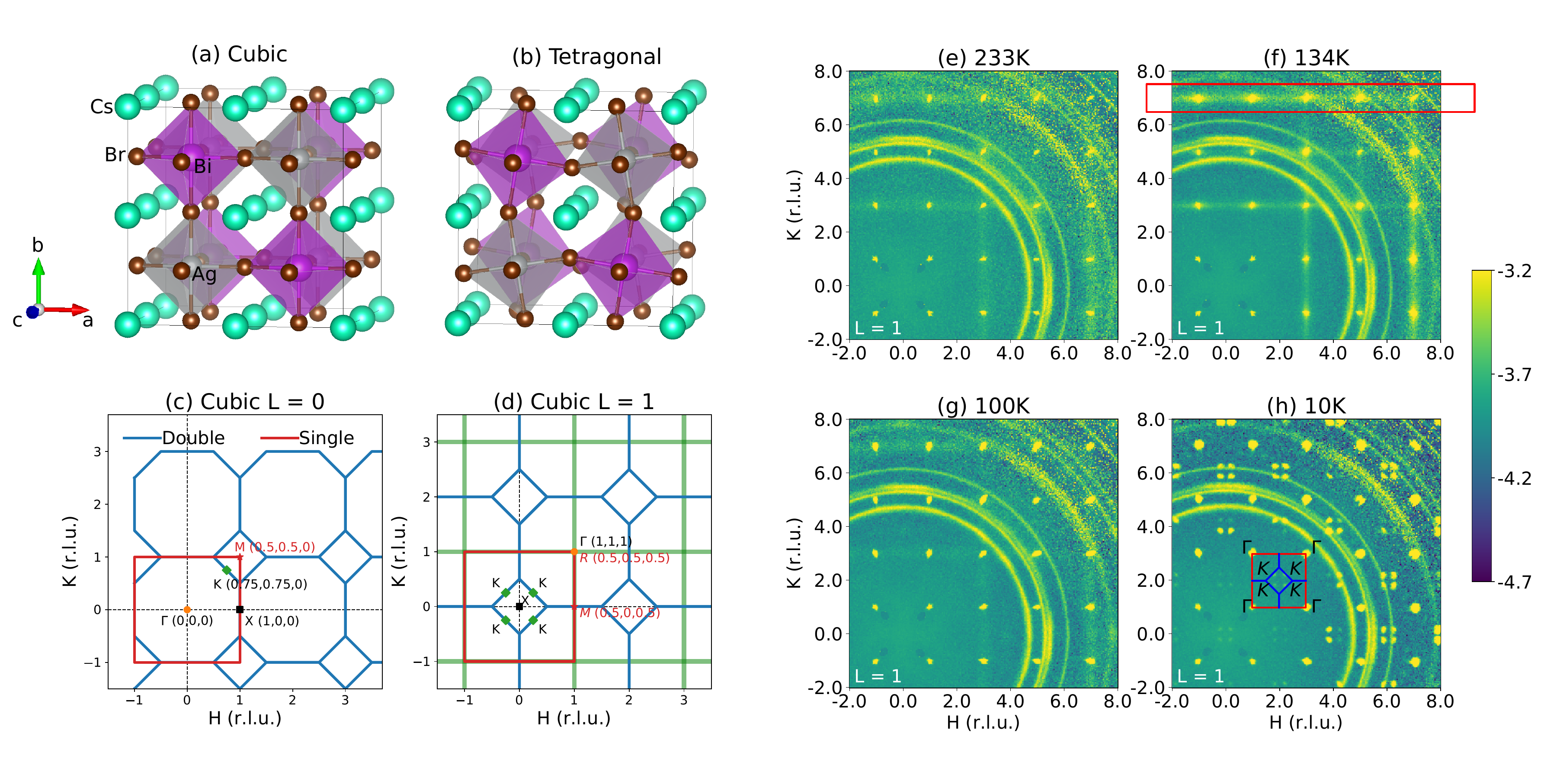}
	\caption{Cs$_2$AgBiBr$_6$ crystal structures, Brillouin zones (BZ), and diffuse scattering. (a,b) Structures of cubic and tetragonal Cs$_2$AgBiBr$_6$. Cs, Ag, Bi and Br atoms are shown in cyan, silver, purple, and brown respectively. The tetragonal rotation axis in (b) is the $c$-axis. (c,d) Projection of cubic structure BZ boundaries in (c) $(L=0)$ plane and (d) $(L=1)$ plane (in double perovskite conventional notation). Blue lines are the double-perovskite BZ boundaries, and red lines are the single perovskite BZ boundaries. Green lines symbolize the location of observed diffuse rods, connecting $\Gamma$ and $X$ points of the double perovskite BZ for $(H,K,L)$ all odd, which corresponds to the BZ edges for the single perovskite notation. Green diamonds represent $K$ points of double-perovskite BZ, which become superlattice peaks at low $T$. (e-h) Neutron diffuse scattering data measured in the reciprocal plane $(L=1)$ using CORELLI ($E$-discriminating cross-correlation turned off) at (e) $T=$233\,K, (f) 134\,K, (g) 100\,K and (h) 10\,K. The $L$-integration range is 0.96 to 1.04 (r.l.u.). The red rectangle in (e) highlights a diffuse rod. Red and blue lines in (h) mark the single-perovskite and double-perovskite BZ edges, respectively. Color maps in (e-h) use a $\log_{10}$ scale. Concentric rings in neutron data are from diffraction by polycrystalline aluminum and copper parts of the sample holder (absent from X-ray measurements).}
	\label{fig:Corelli}
\end{figure*}

Metal halide perovskites (MHPs) exhibit outstanding performance for photovoltaic, optoelectronic and radiation detection applications \cite{huang2017understanding,snaith2018present,stoumpos2013crystal, he2018high,yin2019controlled}, and also attract interest for  thermoelectric \cite{lee2017ultralow, xie2020all} or thermochromic applications \cite{ning2019thermochromic}. 
The efficiency of MHP solar cells has drastically improved over the last decade \cite{jeon2018fluorene,jiang2019surface}, 
enabled by their excellent carrier transport properties, including extremely long carrier lifetimes \cite{herz2018lattice,xing2013long,he2018high,whalley2017perspective,Hoye2018,chen2019imperfections}. 
All-inorganic MHPs  present advantages in terms of stability, and enable the growth of large single-crystals with larger bandgaps well suited for radiation detection or thermochromic applications \cite{wei2019halide,li2018cs2pbi2cl2,ning2019thermochromic,he2021cspbbr}, and their fundamental photophysics could actually be superior to those of hybrid systems \cite{zhang2021all}.
Lead-free MHPs, such as Cs$_2$AgBiBr$_6$, are attracting particular interest because of their lower toxicity and improved stability in air compared to lead-based systems \cite{Lei2021-Cs2AgBiBr6-review,slavney2016bismuth,McClure2016, Luo2018Nature, schade2018structural}. Replacing Pb$^{2+}$ in CsPbBr$_3$ with Ag$^+$ and Bi$^{3+}$ lowers the band gap and improves stability  \cite{Lei2021-Cs2AgBiBr6-review,slavney2016bismuth,filip2016band,zelewski2019revealing}. Further, extremely long photocarrier lifetimes were reported \cite{Lei2021-Cs2AgBiBr6-review,slavney2016bismuth,Hoye2018}. 

Importantly, many studies of inorganic MHPs highlighted the importance of lattice instabilities of the halide octahedra framework, and of strong phonon anharmonicity on the electron-phonon coupling, which is key in the optoelectronic response \cite{yaffe2017local, gehrmann2019dynamic, li2019anharmonicity, lanigan2021two}. Several MHP studies investigated competing low-energy lattice instabilities, resulting from many shallow local minima in the potential energy surface \cite{yang2017spontaneous,marronnier2017structural,marronnier2018anharmonicity,zhao2020polymorphous}, and highlighted the coupling between these instabilities and the electronic states at the edges of the band gap \cite{miyata2017large,yang2017spontaneous,guzelturk2021visualization,zhao2017cu,zhao2020polymorphous,ha2021quasiparticle}. In Cs$_2$AgBiBr$_6$ in particular,  a direct link was proposed between the structural and optical properties and the strong electron-phonon coupling  \cite{schade2018structural,steele2018giant,ning2019thermochromic,zelewski2019revealing,leveillee2021phonon}. 
Moreover, recent studies suggested that atomic vibrations could broaden the conduction band edge and reduce the bandgap in Cs$_2$AgBiBr$_6$ \cite{gebhardt2022electronic}, while enhancing the bandgap in single-halide perovskites \cite{gebhardt2021efficient}.
Yet, a detailed understanding of structural dynamics and phonons in Cs$_2$AgBiBr$_6$ remains elusive owing to a lack of momentum-resolved measurements on single-crystals.

Large amplitude atomic fluctuations in MHPs arise from soft and anharmonic potentials, which also result in very low thermal conductivities and large thermal expansion coefficients  \cite{miyata2017lead,lee2017ultralow,katan2018entropy,gehrmann2019dynamic,bertolotti2017coherent,sakata1980neutron,simoncelli2019unified,haeger2020thermal}.
For instance, Cs$_2$AgBiBr$_6$ is predicted to exhibit ultralow lattice thermal conductivity, which is of fundamental interest and also offers promise for thermoelectric applications \cite{lee2017ultralow,lanigan2021two,haque2018origin,klarbring2020anharmonicity}.  
{Further compounding the need for detailed structural studies, several MHPs were recently predicted to exhibit complex ground states with large unit cells \cite{zhao2020polymorphous}, although this prediction remains to be experimentally confirmed.}
Strongly anharmonic phonons and surprising two-dimensional correlations of Br octahedra tilts were recently revealed in CsPbBr$_3$ \cite{lanigan2021two}, but it remains to be established whether these unusual dynamics also arise in Pb-free MHPs. 
Thus, understanding the atomic structure and dynamics of Cs$_2$AgBiBr$_6$ is critical to rationalize its attractive properties, and to provide insights in the atomic structure and dynamics of the broader MHP family.  

Here, we report the first momentum-resolved investigation of the crystal structure and phonons in Cs$_2$AgBiBr$_6$ single-crystals using  inelastic neutron scattering (INS) and diffuse scattering of both neutrons and x-rays. Our experiments are complemented with first-principles simulations of anharmonic phonons and structural distortions, extended to large simulation scales with a surrogate neural-network potential derived from machine-learning. Our experiments and simulations both reveal a surprisingly complex ground state, and pervasive 2D correlated fluctuations of Br octahedra tilts.

\begin{figure*}
	\centering
	\includegraphics[width=\textwidth]{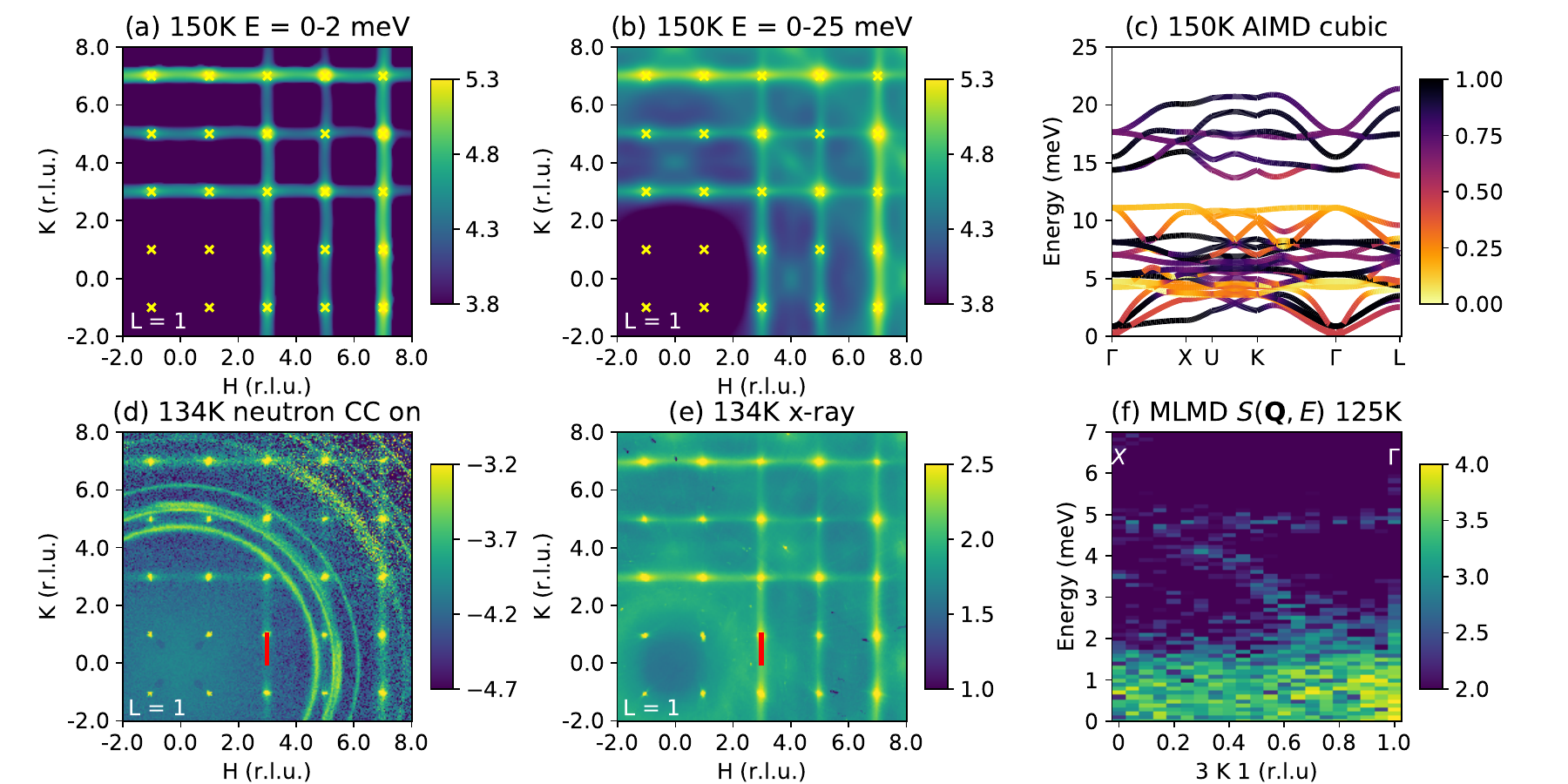}
	\caption{Diffuse rods in cubic phase and damped low-energy optic branch. (a,b) Thermal diffuse scattering (TDS) simulations for cubic phase computed using TDEP 150\,K force-constants. Energy integration ranges 0-2\,meV and 0-25\,meV are shown in (a) and (b) and compared with (d) neutron measurements at 134\,K (with CC turned on) and (e) x-ray measurement at 134\,K. In all TDS simulations, the location of Bragg peaks are indicated by yellow crosses. Panels (c) shows the Br eigenvector weighted phonon dispersions of 150\,K cubic phase. The color maps on dispersions refer to the modulus of Br components in eigenvectors of each mode. As eigenvectors are normalized to one for each mode as $\sum_{j}{e_j^2}=1$, where j is atom and $e_j$ is eigenvector, the minimum of color map is 0 and maximum is 1. (f) $S(\boldsymbol{Q},E)$ along $(3,K,1)$ calculated from 125\,K MLMD trajectory on a 4$\times$20$\times$4 supercell (12800 atoms). The low-energy damped optic branch is responsible to the diffuse rods observed in neutron and x-ray experiment, as marked as red segments in (d) and (e). The color map of panel (c) is in linear scale, while color maps of panel (a-b) and (d-f) are in $\log_{10}$ scale. Concentric rings in neutron data are from diffraction by polycrystalline aluminum and copper parts of the sample holder (absent from X-ray measurements).}
	\label{fig:TDS_cubic}
\end{figure*}

\section{Results and Discussion}
The average structure of Cs$_2$AgBiBr$_6$ is cubic double-perovskite ($Fm\bar{3}m$) at room temperature  \cite{McClure2016,schade2018structural,schade2021crystallographic}. Compared to single halide perovskites ($ABX_3$ prototype, \textit{e.g.} CsPbBr$_3$), the conventional cell of Cs$_2$AgBiBr$_6$ has doubled lattice constants and contains 40 atoms, while the primitive cell of the face centered cubic (FCC) lattice contains 10 atoms. The Ag$^+$ and Bi$^{3+}$ ions form an ordered rock-salt sublattice on the $B$ site of the parent single perovskite. On cooling, Cs$_2$AgBiBr$_6$ undergoes a second-order phase transition at $T_{c1} \simeq 122$\,K  to a tetragonal structure with space group $I4/m$, which was proposed to arise from the freezing of a zone-center soft-mode (out-of-phase Br octahedra rotations about the $c$ axis, $a^0a^0c^-$ in Glazer notation) \cite{schade2018structural,schade2021crystallographic,klarbring2020anharmonicity}. As shown in Fig.~\ref{fig:Corelli}(a,b), the resulting octahedral rotations about the $c$ axis create adjacent layers perpendicular to this axis having opposite phase. 
A recent pair distribution function (PDF) study reported local distortions under high pressure but present even at ambient condition, although the associated time scale was not investigated \cite{girdzis2020revealing}.

\emph{Single-crystal diffuse scattering --}
Neutron diffuse scattering measurements were conducted on a single crystal (mass $\sim 0.2$\,g) as a function of temperature using CORELLI at the Spallation Neutron Source. 
CORELLI uses a wide wavelength band to efficiently map large swaths of wave-vectors $\mathbf{Q}$ in reciprocal space, but also enables the discrimination of purely elastic \textit{vs} inelastic signals with a resolution of $\sim 1\,$meV (for incident neutron energy at 30\,meV), through the use of a semi-random chopper and cross-correlation (CC) technique \cite{ye2018implementation}. 
Large $\mathbf{Q}$-volumes of diffuse x-ray scattering (integrating over the phonon energy spectrum) were also collected with hard x-rays ($E=87.0\,$keV) on Sector 6-ID-D at the Advanced Photon Source (crystal thickness $\sim500\,\mu m$). Sharp Bragg peaks at integer $(H,K,L)$ in both neutron and x-ray experiments indicate the high quality of the crystals and the absence of extra grains.

Figure~\ref{fig:Corelli} illustrates the cubic and tetragonal average crystal structures (a,b), the correspondence between single and double perovskite Brillouin zones (BZ) in (c,d), and shows several diffuse scattering maps extracted from CORELLI data volumes in (e,f,g,h). At all temperatures, weak superstructure peaks are observed at $(H,K,L)$ all odd (double-perovskite notation), equivalent to $R$ points of the single-perovskite BZ, corresponding to the rock-salt ordering of Ag$^+$ and Bi$^{3+}$ ions on the perovskite $B$-site sublattice. Strikingly, in the cubic phase, rods of diffuse scattering intensity are observed in reciprocal planes $L= odd$ in double-perovskite notation, such as $L=1$ in Fig.~\ref{fig:Corelli} (e,f), reminiscent of recent observations in CsPbBr$_3$ \cite{lanigan2021two}. The rod intensity increases upon cooling in the cubic phase, reaching a maximum just above $T_{c1}$, as seen in Fig.~\ref{fig:Corelli} (e,f) 
(see also Supplement Fig.~S3 and S4).
The rods remain faintly visible at 100\,K in the tetragonal phase, see Fig.~\ref{fig:Corelli}(g). While they are observed for CORELLI CC either on or off, the diffuse rods are more intense in the energy-integrated data (CC off) than in the cross-correlated data (nearly elastic scattering), indicating that the rods have an inelastic component, although they also extend into the quasielastic regime, since the cross-correlation energy filtering is approximately $1$\,meV around the elastic channel Fig.~\ref{fig:TDS_cubic}(d). Recent neutron and x-ray scattering measurements on CsPbBr$_3$ revealed similar diffuse rods with an extended quasielastic component over $0\leqslant E\leqslant2\,$meV, for transverse-acoustic (TA) modes dominated by Br motions along the edges of the simple perovskite BZ \cite{lanigan2021two}.
First-principles simulations pointed out their origin in the competition of in-phase \textit{vs} out-of-phase  patterns of collective PbBr$_6$ octahedra rotations corresponding to the lowest-energy TA phonons at the $M$ and $R$ points of the simple perovskite BZ. 

While the behavior observed in Cs$_2$AgBiBr$_6$ establishes a more general character in the MHP family of slow correlated fluctuations first observed in CsPbBr$_3$ \cite{lanigan2021two}, we also note important differences between the two compounds. In particular, the intensity along diffuse rods in cubic CsPbBr$_3$ is clearly modulated with maxima at $M$ and $R$ points \cite{lanigan2021two}, whereas the rods in Cs$_2$AgBiBr$_6$ exhibit no local maximum at $M$ ($X$ point in double perovskite), and the $R$ points correspond to the Ag/Bi ordering superlattice peaks ($\Gamma$ points in double perovskite). The absence of maxima at $M/X$, can be related to the different types of transition into the tetragonal phase. The soft-mode condensation at $\Gamma$ in the double-perovskite
is a second-order transition and yields tetragonal symmetry without introducing superlattice peaks as the unit cell is unchanged. On the other hand, the $M$ point mode in CsPbBr$_3$ condenses at a first-order transition and doubles the unit cell. In addition, the diffuse rods remain strong in the tetragonal phase of CsPbBr$_3$, but become much weaker in tetragonal Cs$_2$AgBiBr$_6$. 
Upon cooling Cs$_2$AgBiBr$_6$ deeper in the tetragonal phase, the $X$ point does not condense, but we show below how a low-lying mode near $K$ points condenses instead  (see Fig.~\ref{fig:Corelli}(h)) and creates a more complex distorted ground state structure than previously reported \cite{McClure2016,schade2018structural,schade2021crystallographic,girdzis2020revealing}.

\begin{figure}
	\centering
	\includegraphics[width=\columnwidth]{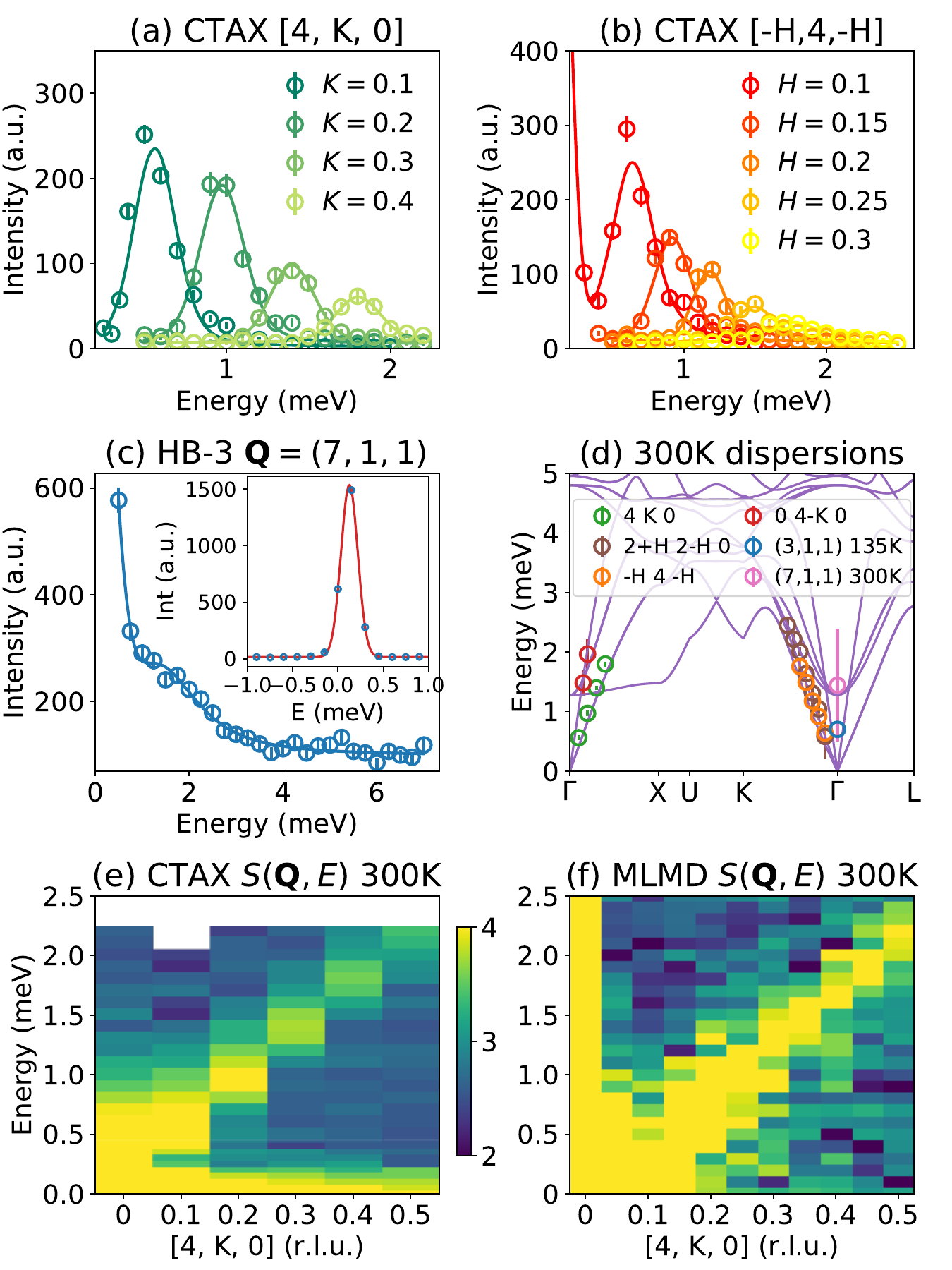}
	\caption{Inelastic neutron scattering spectra and phonon dispersions. (a,b) CTAX measurements along at 300\,K along (a) $[4,K,0]$ and (b) $[-H,4,-H]$. (c) HB-3 measurement at $\boldsymbol{Q}=(7,1,1)$ at 300\,K. The scattered markers in (a-c) are measured data and lines are fitted curves using damped harmonic oscillator models convolving with instrumental resolutions. The inset plot in (c) shows the scan at $\boldsymbol{Q}=(7,1,1)$ across elastic line (markers). The fitted Gaussian curve (red lines) centered at energy 0.126\,meV. This value is used to correct the phonon frequency at zone-center. (d) Renormalized phonon dispersions using cubic phase 300\,K AIMD force-constants compared with experimental data. Markers are fitted phonon energies, and error bars indicate fitted intrinsic phonon linewidths. (e) $S(\boldsymbol{Q},E)$ along $[4,K,0]$ direction measured at CTAX, generated by stacking scans with different $K$, comparing with (f) MLMD calculated $S(\boldsymbol{Q},E)$ on a 4$\times$20$\times$4 supercell (12800 atoms). The color maps of panel (e,f) are in $\log_{10}$ scale.}
	\label{fig:CTAX_dispersions}
\end{figure}

To investigate the phononic origin of the diffuse rods, and anharmonic renormalization in the cubic phase, we performed ab initio molecular dynamics (AIMD) simulations at 300\,K and 150\,K (cubic). The AIMD trajectories were used as input to the temperature-dependent effective potential (TDEP) method to obtain renormalized second-order force-constants and associated phonon dispersions  \cite{hellman2011lattice,hellman2013temperature,hellman2013temperature2}. The renormalized phonon dispersions are shown in Fig.~\ref{fig:TDS_cubic}(c) and Supplement Fig.~S11. As seen on these plots, the $\Gamma-X$ branch is unstable in harmonic calculations but becomes anharmonically stabilized by thermal fluctuations, yielding a low-lying flat branch at $E\sim1.5\,$meV at 300\,K, in good agreement with previous simulations \cite{klarbring2020anharmonicity}, and with our INS measurements of phonon energies shown in Fig.~\ref{fig:CTAX_dispersions}. 
This branch is even softer at 150\,K than at 300\,K and results in more intense thermal diffuse scattering (TDS) (Supplement Fig.~S17), consistent with our measurements showing an increased rod intensity just above $T_{c1} \simeq 122$\,K, since the TDS is proportional to the Bose-Einstein occupation factor and thus highly sensitive to low-energy phonons. We emphasize that the entire $\Gamma-X$ low-energy optic branch softens, instead of just the wave-vector ($\Gamma$) of the soft mode corresponding to the cubic-tetragonal transition. 
The lowest energy optic branch along $\Gamma-X$ is dominated by Br octahedra rotations, as revealed by the phonon dispersions weighted by Br motion amplitudes shown in Fig.~\ref{fig:TDS_cubic} (c) and by phonon animations in the Supplement. The $\Gamma$ mode involves out-of-phase octahedra rotations like the $R$ mode in the single-perovskite, while the $X$  mode involves in-phase  rotations like the $M$ mode in the single-perovskite. 
The dynamical structure factor, $S(\boldsymbol{Q},E)$, was further computed using the renormalized  force-constants in the cubic phase (see Eq.~\ref{eq:SQE} and details in Appendix). Results at 150\,K are shown in Fig.~\ref{fig:TDS_cubic}(a,b). To compare with CORELLI data, the simulated $S(\boldsymbol{Q},E)$ was integrated over $0\leqslant E\leqslant 2$\,meV (quasi-elastic channel) to compare with Fig.~\ref{fig:TDS_cubic}(d), or $0\leqslant E\leqslant 25$\,meV (full TDS corresponding to cross-correlation chopper off or to x-ray diffuse scattering) to compare with Fig.~\ref{fig:TDS_cubic}(e). The excellent agreement between the computed TDS and measured diffuse scattering maps, for both neutron and x-ray scattering, further establishes that the low energy flat branch from Br motions is responsible for the diffuse rods observed in planes $L=odd$. Due to the phase factor, these rods have suppressed intensity in planes $L=even$  (supplement Figs. S1 and S2). 

While our TDS simulations using renormalized force-constants account for the observed diffuse scattering signal very well, the TDEP method still describes the system within an effective second-order harmonic approximation. To fully include the effect of anharmonicity on the phonon spectral functions, without the limitations of perturbation theory, we further used machine-learning based molecular dynamics (MLMD), {with a neural-network force-field trained against AIMD data (see Appendix).} Figure~\ref{fig:TDS_cubic}(f) shows the $S(\boldsymbol{Q},E)$ along $[3,K,1]$ calculated from the 125\,K MLMD trajectory (4$\times$20$\times$4 supercell containing 12800 atoms, see Eq.~\ref{eq:StotMD} and details in Appendix). It clearly reveals the flat low-energy optic branch, which is strongly damped along the entire $\Gamma - X$ segment and results in a broad spectrum with quasielastic component $E\leqslant 2$\,meV. This behavior is a signature of the strong anharmonic damping of Br octahedra rotations. The breakdown of these modes into very low energy fluctuations is responsible for the strong diffuse rods along $\Gamma -X$, as indicated by the red segment along $[3,K,1]$ marked in Fig.~\ref{fig:TDS_cubic} (d,e). {In addition, we performed MLMD on a large 10$\times$10$\times$10 supercell (40000 atoms) to compute the diffuse scattering (see Appendix). The MLMD reproduced the observed rods from Br fluctuations, as shown in supplement Figs. S22, S23, and S24, validating the training of the neural-network force-field.}

\emph{Inelastic neutron scattering  --}
We now compare our INS measurements with our phonon simulations and show that low-energy acoustic modes retain a sharp nature, contrasting strongly with the behavior of soft, broad optic phonons for Br octahedra rotations.
We conducted high-resolution INS on Cs$_2$AgBiBr$_6$ single-crystals using the Cold Neutron Triple-Axis Spectrometer (CTAX) and HB-3 triple-axis spectrometer at the High Flux Isotope Reactor (details in Appendix). 
The TA phonons measured along $[4,K,0]$ and $[-H,4,-H]$ at 300\,K are shown in Fig.~\ref{fig:CTAX_dispersions}(a,b). They exhibit clear peaks with relatively narrow  linewidths.
In contrast, the spectra at $\boldsymbol{Q} = (7,1,1)$ shown in Fig.~\ref{fig:CTAX_dispersions}(c), reveal that the low-energy zone-center optic mode is strongly broadened at 300\,K, with only a broad bump at $\sim 1.5$\,meV. 
Additional CTAX phonon spectra along  $[0,4-K,0]$, $[2+H,2-H,0]$ at 300\,K, and at $(3,1,1)$ at several  temperatures are shown in supplement. 

The INS spectra were fit with a damped harmonic oscillator (DHO) model convolved with the experimental resolution (details in supplement). The resulting phonon energies and linewidths are shown in Fig.~\ref{fig:CTAX_dispersions}(d). As can be seen, our computed renormalized phonon dispersions at 300\,K agree very well with our 300\,K INS measurements. In addition, a clear softening of the zone-center optic mode is observed at 135\,K ($E= 0.70\,$meV) compared to 300\,K ($E= 1.44\,$meV), {confirming that the zone-center soft-mode condensation in the cubic phase yields the tetragonal transition \cite{schade2018structural,schade2021crystallographic,klarbring2020anharmonicity}.}
The phonon spectra measured along $[4,K,0]$ at 300\,K are plotted as a color map in Fig.~\ref{fig:CTAX_dispersions}(e). The sharp nature of acoustic modes is captured well by our MLMD $S(\boldsymbol{Q},E)$ simulations (Fig.~\ref{fig:CTAX_dispersions}(f) and supplementary Fig.~S21), which validates our neural-network force-field, and strongly contrasts with the broad flat optic branch along $\Gamma-X$ shown in Fig.~\ref{fig:TDS_cubic} (f).

\begin{figure*}
	\centering
	\includegraphics[width=\textwidth]{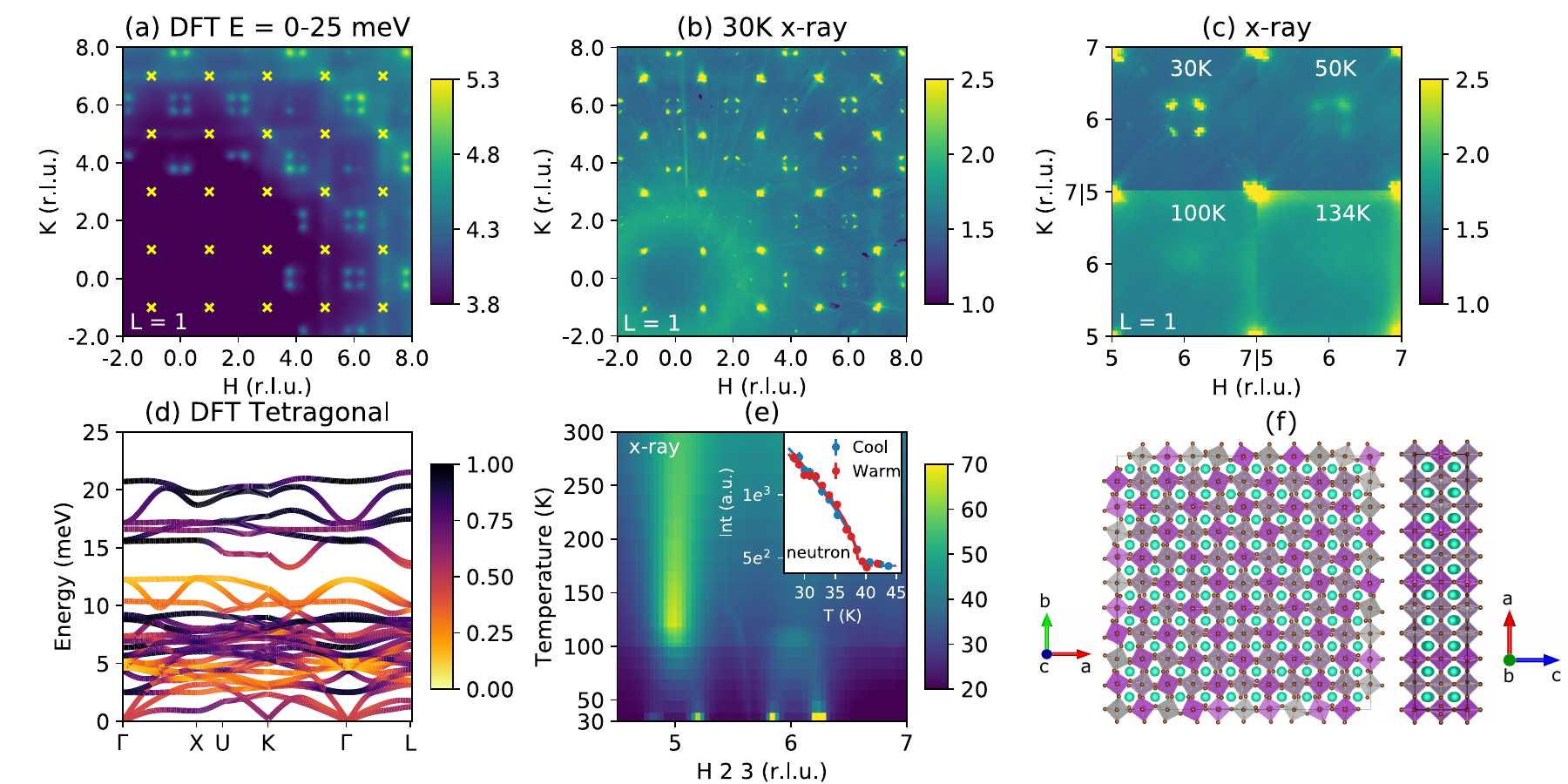}
	\caption{Lattice instability of tetragonal phase at $K$-points and modulated ground state. (a) TDS simulation in tetragonal phase (plane $L=1$) integrated over the energy range 0-25\,meV, computed using interpolated harmonic DFT force-constants. The location of Bragg peaks is indicated by yellow crosses. (b) Experimental x-ray data (plane $L=1$) measured at 30\,K, and zoomed in portion in (c) comparing data at 30, 50, 100, and 134\,K. As seen in (c), $X$ point intensity gradually splits to 4 nearby $K$ points. (d) Phonon dispersions of harmonic DFT tetragonal phase, with color coding indicating the modulus of Br eigenvector. Eigenvectors are normalized to unity for each mode as $\sum_{j}{e_j^2}=1$, where j is atom and $e_j$ is eigenvector, the minimum of the color map is 0 and maximum is 1. (e)  Waterfall plot of x-ray scattering intensity along $[H,2,3]$ (r.l.u) over a large temperature range. The $X$ points at $(5,2,3)$, $(6,2,3)$ start to split $\sim122$\,K on cooling, indicating cubic-tetragonal phase transition. The background discontinuity $\sim90$\,K is due to the change from Helium to Nitrogen cooling gas, while the sudden increase of intensity below 40\,K indicates the phase transition from soft-modes near $K$ points condensing. (f) Predicted exaggerated ground state structure viewed along $c$ and $b$ axis. The color maps of panel (a-c) are in $\log_{10}$ scale, while (d,e) are in linear scale.}
	\label{fig:TDS_tetragonal}
\end{figure*}

\emph{Tetragonal phase instability and ground state --}
As discussed above, the tetragonal phase results from a modulation of the cubic structure by out-of-phase octahedral rotations, due to the soft zone-center optic mode. Indeed, distorting the cubic structure with the eigenvector of the zone-center mode resulted in a tetragonal structure with space group $I4/m$, whose parameters (Table~S1) agree well with reported experimental values \cite{schade2018structural}. 
Further harmonic phonon simulations performed on this structure were stable across most of the BZ, but surprisingly retained a weak instability near the $K$ point,  (0.75,0.75,0) in double perovskite conventional notation, as shown in supplement Fig.~S11 (b). The eigenvector of this unstable mode mainly involves Br and Cs motions and
is illustrated in supplementary animations. 
This phonon instability indicates that the true ground state should exhibit a further distortion and enlarged unit cell. Indeed, the unstable mode matches well the wave-vectors of the superlattice peaks observed in the CORELLI measurements at 10\,K, see Fig.~\ref{fig:Corelli}(h) and low-$T$ x-ray data in Fig.~\ref{fig:TDS_tetragonal}(b,c) (see note \cite{footnote1}). However, we note that superlattice peaks appear slightly away from $K$, closer to $(0.8,0.8,0)$ and are possibly incommensurate. Since the imaginary phonon frequency of the computed soft-mode at $K$ is small, the instability is weak and the phase transition occurs at low temperature. 
As expected, this instability is sensitive to the exchange-correlation functionals and the unit cell volume (supplement Fig.~S13). Barely unstable dispersions were obtained by interpolating force-constants (FC) at different volumes (Fig.~\ref{fig:TDS_tetragonal}(d) and Supplement Fig.~S14), and subsequently were used to compute $S(\boldsymbol{Q},E)$, integrated over $0\leqslant E\leqslant25$\,meV to obtain the TDS in Fig.~\ref{fig:TDS_tetragonal}(a). This can directly be compared with the 30\,K x-ray diffuse scattering in Fig.~\ref{fig:TDS_tetragonal}(b), showing excellent agreement. We note that $K$-points are close to the $X$-points in the BZ (Fig.~\ref{fig:TDS_cubic} (c,d)) and share some similarities in their eigenvectors. Both of the low-energy modes at $K$ and $X$ introduce in-phase octahedral rotations.
However, within $a-b$ layers, instead of having a two single-perovskite periodicity, the soft mode near $K$ modulates the rotation with a larger wavelength. As shown in Fig.~\ref{fig:TDS_tetragonal} (f), the modulation by the eigenvector of mode at (0.8,0.8,0) results in a large $5\times5\times1$ supercell of the original double-perovskite cell, containing 1000 atoms (distortion amplitude exaggerated to help visualize atomic displacements). We see that in addition to Br displacements mainly in the $a-b$ plane, Cs displacements occur along the $c$ axis. The modulation resulting from the incommensurate soft mode near $K$ reduces the symmetry of the tetragonal structure. Our DFT simulations suggest that the ground state could have space-group $Pm$ (\#6) with a $4\times4\times1$ supercell (640 atoms) as modulated by wavevector (0.75,0.75,0), although the monoclinic distortion angle away from $90^{\circ}$ could be very small. If a modulation wave vector of $(0.8,0.8,0)$ is used instead, an even larger $5\times5\times1$ supercell (1000 atoms) results, with a possible space group $Cm$ (\#8) (see details in Supplement Section VIII). We note that distortion energies are exceedingly small (less than 0.1\,meV/atom) and thus sensitive to details of the DFT simulations.

In addition, our low-$T$ x-ray measurements clearly show a broad diffuse intensity maximum at $X$ points at 134\,K (cubic phase) and strengthening at 100\,K (tetragonal) and gradually splitting into sharper maxima close to $K$ points, eventually condensing into intense superstructure Bragg spots (Fig.~\ref{fig:TDS_tetragonal}(b,c)). A gradual temperature evolution is revealed in Fig.~\ref{fig:TDS_tetragonal}(e), which shows the intensity along $(H,2,3)$ from 300\,K to 30\,K. 
Cooling below $T_{c1}=122\,$K, each $X$-point maximum ($H=\{5,6\}$) splits into two maxima around $K$ points ($H= \{4.75, 5.25\}$ and $\{5.75,6.25\}$ respectively). Note that the path in Fig.~\ref{fig:TDS_tetragonal}(e) is integrated over a wide band to include $K$ points at $(H,2,3\pm0.25)$. A clear increase in intensity near $K$ occurs below 40\,K, indicating the transition to the distorted ground state.
This is entirely consistent with the condensation of the soft mode near $K$ and its competition with the $X$ point near-instability  (see discussion in Supplement Sections VI and VII).
Using CTAX, we tracked the superlattice peak intensity 
over a cooling and warming cycle (details in Supplement Section VII), as shown in the inset of Fig.~\ref{fig:TDS_tetragonal}(e). The intensity was fit to $A\cdot(T_{c2}-T)^x+b$, where $A$, $T_{c2}$, $x$ and $b$ are parameters and $T$ is temperature. The fit results for $T_{c2}$ from cooling and warming curves are $39.36\pm0.34$ and $38.78\pm0.20$\,K, respectively. The coincidence of cooling and warming curves and $T_{c2}$ values supports that this phase transition is of second order (or weak first order), and originates from the condensation of a soft mode near the $K$-point. 
We note that during manuscript preparation, a phase transition at $T\sim$38\,K was reported based on Raman spectroscopy, supporting our findings, although that study did not discuss the ground state structure \cite{cohendiverging2022}.

\section{Conclusion}

In summary, we reported the first momentum-resolved investigations of complex dynamic fluctuations in the double-perovskite Cs$_2$AgBiBr$_6$ based on single-crystal neutron and x-ray scattering measurements, rationalized with first-principles simulations augmented with large scale machine-learning models. We revealed the occurrence of systematic diffuse scattering rods arising from slow, correlated 2D fluctuations of Br octahedra derived from strongly anharmonic low-energy optic phonons. While arising from distinct phonon branches, these large fluctuations share similarities with the overdamped zone-boundary acoustic modes in the single-perovskite CsPbBr$_3$, suggesting a general behavior of slow correlated 2D fluctuations in MHPs. Critically, a new low-symmetry phase was discovered at $T\leqslant T_{c2}=39$\,K, emerging from a phonon instability of the tetragonal phase near the $K$-point, and resulting in an unusually complex ground state with several hundred atoms in the unit cell. Our results suggest ubiquitous lattice instabilities in the halide perovskite family resulting from their anharmonic potential energy surface, and reveal a much more complex ground state than previously known in Cs$_2$AgBiBr$_6$. The complex anharmonic lattice dynamics yield large thermal fluctuations and strongly impact the electron-phonon coupling and thus the thermal and optoelectronic properties of these compounds.

\section*{Acknowledgements}
We thank Olle Hellman for providing access to the TDEP software package. XH, MKG and OD were supported by the U.S. Department of Energy, Office of Science, Basic Energy Sciences, Materials Sciences and Engineering Division, under Award No. DE-SC0019978. TLA was supported by the U.S. Department of Energy, Office of Science, Basic Energy Sciences, Materials Sciences and Engineering Division, under Award No. DE-SC0019299.  
Work at the Materials Science Division at Argonne National Laboratory (characterization, x-ray and neutron scattering measurements and analysis) was supported by the U.S. Department of Energy, Office of Science, Office of Basic Energy Sciences, Materials Sciences and Engineering Division.
The synthesis of materials is based upon work supported by the Department of Energy / National Nuclear Security Administration under Award Number(s) DE-NA0003921. 
This research used resources at the High Flux Isotope Reactor and Spallation Neutron Source, both DOE Office of Science User Facilities operated by the Oak Ridge National Laboratory. 
This research used resources of the Advanced Photon Source, a U.S. Department of Energy (DOE) Office of Science User Facility, operated for the DOE Office of Science by Argonne National Laboratory under Contract No. DE-AC02-06CH11357.
Theoretical calculations were performed using resources of the National Energy Research Scientific Computing Center, a U. S. DOE Office of Science User Facility supported by the Office of Science of the U. S. Department of Energy under Contract No. DE-AC02-05CH11231.

\section*{Appendix A: Sample Synthesis}
48\,\% w/w aq. Hydrobromic acid (HBr) and Bismuth (III) bromide (BiBr$_3$) (metals basis) with purity of 99\,\% were purchased from Alfa Aesar, and Silver bromide (AgBr) (metals basis) with purity of 99\,\% and Cesium Bromide (CsBr) with 99.9\,\% purity were purchased from Sigma Aldrich. All the chemicals were directly used as received without any further purification process. Briefly, Cs$_2$AgBiBr$_6$ was synthesized through a solution-processed method with some modification \cite{slavney2016bismuth}. 4.85\,g CsBr, 2.15\,g AgBr, and 5.11\,g BiBr$_3$ were dissolved into 125\,mL HBr solution in a sealed container at 120\,$\degree$C on a hot plate after about two hours stirring. The precursor solution was filtered at 120\,$\degree$C before single crystal growth. Finally, large single crystals can be grown by gradually reducing the heating temperature of hot plate to 80\,$\degree$C at a rate of 0.5\,$\degree$C/h. To avoid any light induced chemicals decomposition, the Cs$_2$AgBiBr$_6$ single crystals were grown in the dark condition. A photograph of a  Cs$_2$AgBiBr$_6$ single-crystal is shown in Supplement.

\section*{Appendix B: X-ray and neutron scattering}

\subsection*{Diffuse X-ray scattering}
Diffuse X-ray scattering data were collected at the Advanced Photon Source on sector 6-ID-D using an incident energy of 87.0 keV. Sample temperatures from 30\,K to 300\,K were controlled with He and N$_2$ gas flow. During the measurements, the samples were continuously rotated about an axis perpendicular to the beam at 1$^{\circ}s^{-1}$ over 370$^{\circ}$, with images recorded every 0.1\,s on a Dectris Pilatus 2M detector with a 1-mm-thick CdTe sensor layer. Three sets of rotation images were collected for each sample at each temperature to fill in gaps between the detector chips. The resulting images were stacked into a three-dimensional array, oriented using an automated peak search algorithm and transformed in reciprocal space coordinates, allowing S$(\mathbf{Q})$ to be determined over a range of $\sim\pm$15 $\AA^{-1}$ in all directions. The x-ray scattering intensities are proportional to S(\textbf{Q}), weighted by the square of the atomic scattering factors for each element. Further details are given in Ref.~\cite{Krogstad2020diffuse}.

\subsection*{Diffuse neutron scattering}
Neutron diffuse scattering experiments were conducted using the CORELLI instrument at SNS, Oak Ridge National Laboratory (ORNL) as a function of temperature. Unlike conventional single-crystal x-ray or neutron diffractometers, CORELLI uses a wide wavelength band to cover large swaths of wave-vectors $\mathbf{Q}$ in reciprocal space, but enables the discrimination of purely elastic \textit{vs} inelastic signals with a resolution of $\sim 1\,$meV, through the use of a semi-random chopper and cross-correlation technique. The crystal was mounted on an aluminum pin with thin copper wire, with the pin wrapped in Gd-foil to reduce background, and the assembly was loaded on closed-cycle refrigerator. Measurements were performed on cooling down to $T\sim10\,$K as well as warming to $\sim400\,$K. At selected temperatures, the sample was rotated over a wide range of angle ($360^{\circ}$) to collect large continuous data volumes in reciprocal space and investigate the diffuse scattering.

\subsection*{Inelastic neutron scattering}
Inelastic neutron scattering (INS) measurements were conducted on the Cold Neutron Triple-Axis Spectrometer (CTAX) and HB-3 triple-axis spectrometer at the High Flux Isotope Reactor (HFIR), ORNL. A few single crystals of Cs$_2$AgBiBr$_6$ were used with mass 0.2$\sim$0.4\.g. Samples had mosaics of about 30$^\prime$.
Using CTAX, we measured acoustic modes along $[4,K,0]$ at 300\,K and 100\,K, $[-H,4,-H]$ at 300\,K, $[0,4-K,0]$ at 300\,K, $[2+H,2-H,0]$ at 300\,K, and optic mode at $\boldsymbol{Q} = (3,1,1)$ at 550\,K, 300\,K, and 135K. Using HB-3, we measured the lowest optic mode at $\boldsymbol{Q}=(7,1,1)$ at 300\,K.

In CTAX experiments, the fixed final energy mode was used with $E_f = 5$\,meV for measurement along $[4,K,0]$ at 300\,K and 100\,K, and with $E_f=4.8$\,meV for measurements along $[-H,4,-H]$ at 300\,K, $[0,4-K,0]$ at 300\,K, $[2+H,2-H,0]$ at 300\,K and $\boldsymbol{Q} = (3,1,1)$ at 550\,K, 330\,K and 135\,K. PG002 monochromators and analyzers were used. The collimation settings were 30$^\prime$-$100^\prime-80^\prime-120^\prime$, and energy resolution was $\sim0.3$\,meV at zero energy transfer. The high-order energy contamination of neutron beam was removed by a cooled Be filter. Samples were mounted in the $(H,K,0)$ plane using Bragg peaks $(4,0,0)$ and $(0,4,0)$ for measurements along $[4,K,0]$ and $[2+H,2-H,0]$. The scattering plane $(3H,K,H)$ defined by $(3,-1,1)$ and $(3,1,1)$ was used for measurements at $\boldsymbol{Q} = (3,1,1)$, along $[-H,4,-H]$, and $[0,4-K,0]$. While $[-H,4,-H]$ is not within the $(3H,K,H)$ scattering plane, the required tilts were reachable with the instrument goniometer. We note that $[4,K,0]$ and $[0,4-K,0]$ both correspond to wave vectors along $\langle001\rangle$, and are TA modes and longitudinal acoustic (LA) modes from $\Gamma$ to $X$, respectively. Along $[2+H,2-H,0]$ and $[-H,4,-H$, we probe 
the two non-degenerate TA modes from $\Gamma$ to $K$, which have distinct polarizations. 

In the HB-3 experiment, the fixed final energy mode was used with $E_f=13.5$\,meV. PG002 monochromators and analyzers were used. The collimation setting was 30$^\prime$-$40^\prime-40^\prime-70^\prime$, and the energy resolution was $\sim1$\,meV at zero energy transfer. The sample was mounted in the $(H,K,0)$ plane. $\boldsymbol{Q} = (7,1,1)$ was reached with tilts of the instrument goniometer. 

\section*{Appendix C: First-Principles Simulations}

First-principles phonon simulations were performed using VASP \cite{kresse1993ab,kresse1994ab,kresse1996efficiency,kresse1996efficient} in the cubic and tetragonal phases of Cs$_2$AgBiBr$_6$, with a special attention paid to anharmonic effects. We first describe the results of phonon simulations within the small-displacement approach and the harmonic approximation, as implemented in Phonopy \cite{phonopy}, in the cubic phase ($Fm\overline{3}m$, space group 225), using a $2\times2\times2$ supercell of the double-perovskite cell with experimental lattice constant 11.2695 $\AA$ from ref \cite{ji2020lead} (details in Supplement). These harmonic simulations result in unstable phonon dispersions, as shown in Supplement Fig.~S8 and Fig.~S9. The unstable mode at the $\Gamma$ point is triple-degenerate, corresponding to out-of-phase rotations of Br octahedra about the three $\langle 100 \rangle_{\rm cubic}$ axes.  Modulating the cubic structure according to the eigenvector of one of these unstable modes results in the tetragonal structure. The structure was then relaxed using VASP until the force on each atom was smaller than $10^{-4}$\,eV$/\AA$, yielding the expected $I4/m$ symmetry (space group number 87), see details in Supplement. The relaxed structure has lattice constants $a = b = 11.162\,\AA$, $c = 11.488\,\AA$, which compare well with experimental values $a = 7.8794\times\sqrt2=11.143\ \AA$, $c = 11.324\,\AA$ \cite{schade2018structural}. After relaxing the tetragonal structure, further harmonic phonon simulations were performed. The phonon dispersions were then nearly stable, with only a weak instability around the $K$ point ((0.75,0.75,0) in double perovskite conventional notation) as shown in Supplement Fig.~S8. This instability indicates that the true ground state exhibits a further modulated distortion and enlarged unit cell. Indeed, the location of the instability around $K$ matches well the wave-vectors of the superlattice peaks observed in the neutron and x-ray diffuse scattering measurements below 40\,K. Since the imaginary phonon frequency around this $K$ point mode is small, the instability is weak and the  phase transition happens at low temperature, below $40$\,K.

The dynamical structure factor $S(\mathbf{Q},E)$ was computed for the cubic and tetragonal phases using either the harmonic or renormalized force-constants to obtain phonon frequencies $\omega_s$ and eigenvectors $e_{ds}$.
The simulated phonon intensity was calculated as \cite{squires1996introduction}:
\begin{eqnarray} 
& S({\bf Q},E) \propto \nonumber  \\ 
& \sum\limits_{s, \mathbf{\tau}} \frac{1}{\omega_{s}}  \left| \sum\limits_{d} \frac{\overline{b_d}}{\sqrt{M_d}} \rm{exp}(-W_d)\rm{exp}(i\mathbf{Q}\cdot\mathbf{d})(\mathbf{Q}\cdot\mathbf{e}_{ds})\right|^2 \nonumber \\ 
& \times \langle n_s +1\rangle \delta(\omega - \omega_s)\delta(\mathbf{Q}-\mathbf{q} -\mathbf{\tau})
\label{eq:SQE}
\end{eqnarray}
where $\overline{b_d}$ is neutron scattering length of atom $d$, $\boldsymbol{Q} = \boldsymbol{k} - \boldsymbol{k^{\prime}}$ is the wave vector transfer, and $\boldsymbol{k^{\prime}}$ and $\boldsymbol{k}$ are the final and incident wave vector of the scattered particle, $\textbf{q}$ the phonon wave vector, $\omega_s$ the eigenvalue of the phonon corresponding to the branch index $s$, $\bm \tau$ is the reciprocal lattice vector, $d$ the atom index in the unit cell, $\exp(-2W_d)$ the corresponding Debye-Waller factor, and $n_s = \left[\exp\left(\frac{\hbar\omega_s}{k_{\rm B}T}\right)-1\right]^{-1}$ is the mean Bose-Einstein occupation factor.
The phonon eigenvalues and eigenvectors in Eq.~\ref{eq:SQE} were obtained by solving dynamical matrix using Phonopy~\cite{phonopy}. The thermal diffuse scattering (TDS) signals are obtained by integrating $S(\boldsymbol{Q},E)$ over energy.

\section*{Appendix D: Machine-Learning augmented Molecular Dynamics (MLMD)}
We have extended the molecular dynamics simulation to a larger scale to access the well-resolved dynamics in ($\boldsymbol{Q}$,\,$E$) space using a machine-learned force-field (MLFF) based on a neural-network potential. To generate a force-field that describes the dynamics over a broad temperature range, we generated AIMD training datasets over a range of temperatures using VASP \cite{kresse1993ab,kresse1994ab,kresse1996efficiency,kresse1996efficient}. These simulations were performed with the  strongly constrained and appropriately normalized (SCAN) exchange-correlation functional \cite{sun2016accurate}. We performed AIMD on a 40-atom unit cell, with a plane wave kinetic energy cutoff of 400\,eV and a 2$\times$2$\times$2 Monkhorst-Pack grid of $k$-points \cite{monkhorst1976special}. The self-consistent convergence threshold for electronic minimization was set to $10^{-6}$ eV. We ran simulations from 100\,K to 900\,K in 100\,K intervals, with each trajectory lasting about 3-5\,ps and a time step of 2\,fs. All AIMD simulations used the 300\,K experimental lattice constant 11.27\,$\AA$ \cite{schade2018structural}. The temperature of the system was controlled by a Nos\'e-Hoover thermostat. We then used the DEEPMD \cite{wang2018deepmd} code to generate a machine-learned force-field (MLFF) based on a neural network, which reproduces as best as possible the AIMD dynamics. The generated MLFF was used to compute the pair distribution function and mean squared displacements and checked against AIMD results. The LAMMPS package was then used for MLMD classical simulations with the trained MLFF \cite{plimpton1995fast}. 
We calculated the $S(\mathbf{Q},E)$ from MLMD trajectories at 125\,K and 300\,K along $[4,K,0]$ and $[3,K,1]$. For these simulations, we used large supercells (4$\times$20$\times$4 supercell containing 12800 atoms), and computed the atomic trajectories up to $\sim100$\,ps (with a time step of 1\,fs) within an NVT ensemble. The simulation at 125\,K and 300\,K used experimental lattice constants 11.23 and 11.27\,$\AA$ respectively \cite{schade2018structural}. These extensive simulations provide momentum and energy resolutions of $\sim0.1\,\mathrm{\AA^{-1}}$ and $\sim$0.1\,meV, respectively, enabling us to resolve the dispersions. In neutron scattering experiments, the scattering intensity has contributions from both the coherent dynamical structure factor ($S_\mathrm{{coh}}^{IJ}$($\boldsymbol{Q}$, $E$)) attributed to correlated dynamics between $I^\mathrm{{th}}$ and $J^\mathrm{{th}}$ element, and incoherent dynamical structure factor ($S_\mathrm{{inc}}^{I}$($\boldsymbol{Q}$, $E$)) from  uncorrelated self dynamics of the $I^\mathrm{{th}}$ element, respectively defined as:
	\begin{equation}
	S_{\mathrm{coh}}^{IJ}(\boldsymbol{Q},E)=\frac{1}{2\pi}\int_{0}^{\infty}\sum_{i,j} e^{-i\boldsymbol{Q}\cdot(R_i^I(0)-R_j^J(t))}e^{iEt/\hslash}dt
	\end{equation}
	\begin{equation}
	S_{\mathrm{inc}}^{I}(\boldsymbol{Q},E)=\frac{1}{2\pi}\int_{0}^{\infty}\sum_{i} e^{-i\boldsymbol{Q}\cdot(R_i^I(0)-R_i^I(t))}e^{iEt/\hslash}dt
	\end{equation}
\noindent where $R_i^I$($t$) is the $i^{th}$ atom position of element type $I$ at time $t$, and the summation indices $i$ and $j$ represent the sum over $I^{th}$ and $J^{th}$ kind of elements in the simulation cell.
The total measured neutron scattering intensity is then proportional to:
\begin{equation}
	S_{\mathrm{tot}}(\boldsymbol{Q},E) \propto \sum_{I,J} b_{\mathrm{coh}}^I b_{\mathrm{coh}}^J \, S_{\mathrm{coh}}^{IJ}(\boldsymbol{Q},E)+ \sum_{J} (b_{\mathrm{inc}}^J)^2\, S_{\mathrm{inc}}^J(\boldsymbol{Q},E)
\label{eq:StotMD}
\end{equation}
where $b_{\mathrm{coh}}^I$ and $b_{\mathrm{inc}}^I$ are the coherent and incoherent scattering length of $I^{\mathrm{th}}$ element and summation index $I$ and $J$ runs over different elements in the unit cell.

Diffuse scattering intensity calculations were performed with MLMD using a 10$\times$10$\times$10 supercell containing 40000 atoms. We ran MLMD simulations at 125\,K and 300\,K with durations of about 100\,ps with 1\,fs time steps. We computed the total scattering intensity, $S$($\boldsymbol{Q}$) at 0.1\,ps interval ($\sim$1000 configurations) according to:
\begin{equation}
	S(\boldsymbol{Q}) \propto \int_{0}^{\infty} \lvert \sum_{I} b_{\mathrm{coh}}^I \sum_{i \in I}e^{-i\boldsymbol{Q}\cdot R_i(t)}\rvert ^2dt	
\end{equation}
where $i$ is an index running over all atoms of type $I$, and $b_{\mathrm{coh}}^I$ is the coherent neutron scattering length of atom type $I$.  
The diffuse scattering intensity of pair $I-I$ is defined as:
\begin{equation}
	S^{I-I}(\boldsymbol{Q}) \propto (b_{\mathrm{coh}}^I)^2 \int_{0}^{\infty} \lvert \sum_{i \in I} e^{-i\boldsymbol{Q}\cdot R_i(t)}\rvert ^2dt
\end{equation}
where index $i$ enumerates atoms of element type $I$. Note $S(\boldsymbol{Q})\neq \sum\limits_{I} S^{I-I}(\boldsymbol{Q})$.
\newline
\par
The $S$($\boldsymbol{Q}$) and $S^{I-I}(\boldsymbol{Q})$ computed at different time points were then averaged to obtain the diffuse scattering pattern.

\newpage

\bibliography{references-Cs2AgBiBr6}

\begin{thebibliography}{10}
\bibitem{footnote1}
We note that during the cubic-tetragonal phase transition, the unit cell does
  not change except for the elongated ratio $c/a>1$. As a result, the BZ is
  nearly the same in tetragonal and cubic phases but with the reciprocal
  $c^{*}$ parameter shortened. Thus, we keep the same notation for high
  symmetry BZ points and $K$ points refer to (0.75,0.75,0) in planes $L={\rm
  integer}$.
\end{thebibliography}
\bibliographystyle{unsrt}

\end{document}